\documentclass[preprint,12pt,authoryear]{elsarticle}

\usepackage{float}      
\usepackage{graphicx}   
\usepackage{pdflscape}  
\usepackage{afterpage}  
\usepackage{amssymb}    
\usepackage{amsthm}     
\usepackage{amsmath}    
\usepackage{algorithm}  
\usepackage{algpseudocode} 
\usepackage{placeins}      

\usepackage{booktabs}   
\usepackage{rotating}   
\usepackage{subfigure}  
\usepackage{threeparttable}             
\usepackage[margin=0.85in]{geometry}    
\usepackage[skip=0.6\baselineskip]{caption} 
\usepackage{array}      
\newcolumntype{L}[1]{>{\raggedright\let\newline\\\arraybackslash\hspace{0pt}}m{#1}}
\newcolumntype{C}[1]{>{\centering\let\newline\\\arraybackslash\hspace{0pt}}m{#1}}
\usepackage[dvipsnames]{xcolor}
\usepackage{hyperref}                   
\hypersetup{colorlinks=true}
\usepackage{comment}

\journal{SUMO Conference 2026}

\begin{document}

\begin{frontmatter}

\title{Revisiting mesoscopic traffic flow simulation in SUMO: Limitations, analysis, and an alternative}

\author[a]{Ying-Chuan Ni\corref{cor1}} \ead{ying-chuan.ni@ivt.baug.ethz.ch}
\author[a]{Alina Akopian}
\author[a]{Anastasios Kouvelas}
\author[a]{Michail A. Makridis}

\affiliation[a]{organization={Traffic Engineering Group, Institute for Transport Planning and Systems, ETH Zurich},
            addressline={Stefano-Franscini-Platz 5}, 
            city={Zürich},
            postcode={8093}, 
            country={Switzerland}}

\cortext[cor1]{Corresponding author}

\begin{abstract}
Mesoscopic traffic flow models combines the merits of both macroscopic and microscopic models by capturing individual vehicle behavior in great detail and remaining the computational efficiency. At the time of this study, the mesoscopic model proposed by \citet{Eissfeldt2004Modelling} is used in Simulation of Urban MObility (SUMO). The movement of vehicles is governed by dynamic headways between edges. However, the model does not fully comply with the principle of the Lighthill–Whitham–Richards (LWR) model. Several problems are identified, including the incomplete consideration of queue dynamics and the limited implementation of backward traveling spaces. Two case study scenarios demonstrate that the problems lead to unrealistic onset and recovery pattern of congestion. The magnitude of congestion is generally underestimated with this model. To address these drawbacks, a proper mesoscopic discrete-time implementation of link transmission model, which follows the LWR principle, is proposed. By explicitly incorporating backward traveling spaces to capture queue spillback phenomena, the proposed model provides a more precise representation of congestion dynamics. The link density outputs are consistent with the kinematic wave theory and the microscopic traffic simulation in SUMO, thus verifying its theoretical accuracy.
\end{abstract}

 
\begin{keyword}
Backward traveling space \sep Congestion propagation \sep Link transmission model \sep Mesoscopic traffic simulation \sep Queue spillback \sep SUMO
\end{keyword}

\end{frontmatter}

\section{Background}
Mesoscopic traffic simulation tracks individual vehicles while preserving key properties of macroscopic traffic flow. Compared to microscopic simulation, it requires substantially lower computational cost as detailed vehicle-to-vehicle interactions are not explicitly modeled. In the meanwhile, it offers finer-level details that macroscopic traffic modeling is unable to consider, such as behavioral heterogeneity and first-in-first-out condition for multi-commodity flow that has multiple turning directions on certain roads.

Several open-source tools, including DynaMIT~\citep{Ben-Akiva2012ANetworks} and MATSim~\citep{Axhausen2016TheMATSim}, have implemented mesoscopic traffic simulation function. However, many problems with regard to the simulated traffic dynamics in these tools were pointed out in \citet{Ni2026SimulatingNetworks}.

Simulation of Urban MObility (SUMO), an open-source traffic simulation tool \citep{Lopez2018MicroscopicSUMO}, also has a mesoscopic traffic simulation function, which was designed following the model proposed by \citet{Eissfeldt2004Modelling} at the time of this work. The function, hereinafter referred to as MESO, has been adopted for the simulation of large-scale road traffic networks. For instance, \citet{Ambuhl2023UnderstandingDiagram} used MESO to simulate congestion propagation in road networks and investigated the correlation between network traffic collapse and percolation of congestion clusters.

However, \citet{Ni2026SimulatingNetworks} has shown that MESO exhibits several deficiencies that are particularly evident when simulating interrupted traffic flow in urban road networks. One of the fundamental issues pertains to the problematic integration of concepts in cell transmission model (CTM) and link transmission model (LTM). This work aims to provide an in-depth analysis of the problems in the original model \citep{Eissfeldt2004Modelling} and to present an alternative that is consistent with microscopic simulation with a case study.

\section{Recap of Eissfeldt's queue dynamics}
Eissfeldt's model discretizes a link into multiple segments and introduces a state of each segment, which can be either free or jammed. The first vehicle is ready to exit the segment after the free-flow travel time of the segment. This section provides a recap of the modeling logic. The original description can be found in \citet{Eissfeldt2006SimulatingQueues}.

\begin{figure}[!ht]
  \centering
  \includegraphics[width=\textwidth]{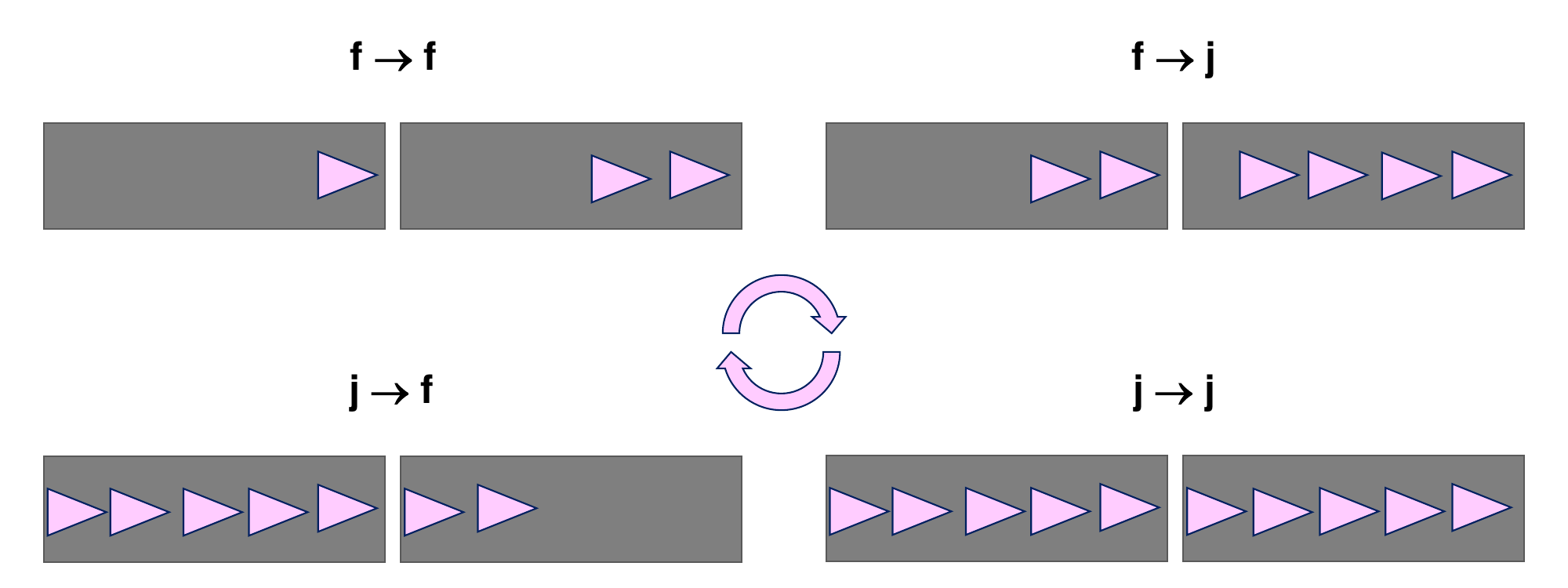}
  \caption{Replica of the sequence of possible queue states in \citet{Eissfeldt2006SimulatingQueues}}\label{fig:Eissfeldt}
\end{figure}

Figure~\ref{fig:Eissfeldt} illustrates the possible evolution of states between two segments on a motorway segment. For each condition, a segment-to-segment headway $\tau$ is applied. Given that the saturation flow is $q^{\max}$, when going from a free segment to another free segment, the minimum headway $\tau_{\text{ff}}=1/q^{\max}$ is used as the outflow is unconstrained. When a vehicle queue starts to build up in the downstream segment, the article described that the minimum headway remains $\tau_{\text{fj}} = \tau_{\text{ff}}$ unless the queue length in the downstream segment reaches its maximum $N$, when a small increase of $\tau_{\text{fj}}$ might occur yet ignored.

The traffic may continue to evolve to the next condition that both segments are fully jammed. In this case, the vehicle in the upstream segment needs to wait until a vacancy travels upstream through the downstream segment when a vehicle exit the downstream segment. Therefore, the waiting time is calculated as $\delta t = \tau_{\text{jj}} \cdot N$. Afterward, the jammed queue in the downstream segment starts to resolve. The headway of a vehicle leaving a jammed segment $\tau_{\text{jf}}$ is usually larger than the minimum headway $\tau_{\text{ff}}$, which is the reason of widening traffic jam on a motorway. This is similar to the concept of capacity drop in macroscopic traffic flow theory \citep{Han2025CapacityData}. The model performance was partially validated with speed data on a motorway segment.

Accordingly, MESO implemented this logic in a discrete-time manner. At every simulation step for every segment, whether the first vehicle can exit or not is checked based on the segment-to-segment conditions explained above.

\section{Problems}
\label{sec:problem}
This section first discusses the fundamental problems identified in Eissfeldt's model and then points out how they affect the current MESO function.

\subsection{Similarity between CTM and LTM}
The distinction between the free and jam states for a segment is originally a concept applied in CTM together with the assumption that the traffic state is always homogeneous within a cell \citep{Daganzo1994TheTheory}. The critical density on a triangular fundamental diagram (FD) is usually used to distinguish free and jam traffic states. By knowing the density or number of vehicles in the cell, the outflow and inflow can be determined through a sending--receiving mechanism. 

In Eissfeldt's model, a distinction logic is also used to determine the headway between segments. However, vehicle inflow and outflow are instead meant to be determined by the free-flow travel time and backward space travel time, which are actually concepts applied in LTM \citep{Yperman2007TheLoading,Yperman2007ADelays}, the discretization of Newell's formulation~\citep{Newell1993ATheory} of the Lighthill-Whitham-Richards (LWR) model \citep{Lighthill1955OnRoads,Richards1956ShockHighway}. Different from CTM, LTM tackles the heterogeneous within-link traffic dynamics. \citet{Osorio2011DynamicDistributions} provided a comprehensible way to interpret the physics of LTM with a two-queue format on a road link.

The mixture of concepts from two models causes the model to produce unrealistic traffic dynamics. In particular, the distinction between free and jam segments limits the ability of the model to simulate backward traveling spaces, which will be discussed in subsection~\ref{subsec:invalid}.

\subsection{Invaid consideration of queue dynamics} \label{subsec:invalid}
\begin{figure}[!ht]
  \centering
  \includegraphics[width=0.525\textwidth]{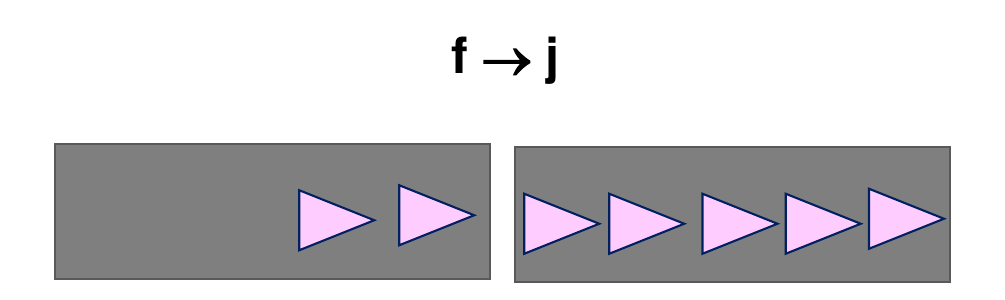}
  \caption{Queue states in the free-jam condition}\label{fig:incomplete}
\end{figure}

For the free-jam condition, as shown in Figure~\ref{fig:incomplete}, the validity of the logic for headway $\tau_{\text{fj}}$ is questioned. Theoretically, the segment-to-segment headway is not affected by the state of the upstream segment but largely depends on whether there is an available space in the downstream segment. The first vehicle in the upstream segment should only be able to exit when a space, or vacancy, travels through the downstream segment, which is the logic currently applied only in the jam-jam condition.

\begin{figure}[!ht]
  \centering
  \includegraphics[width=\textwidth]{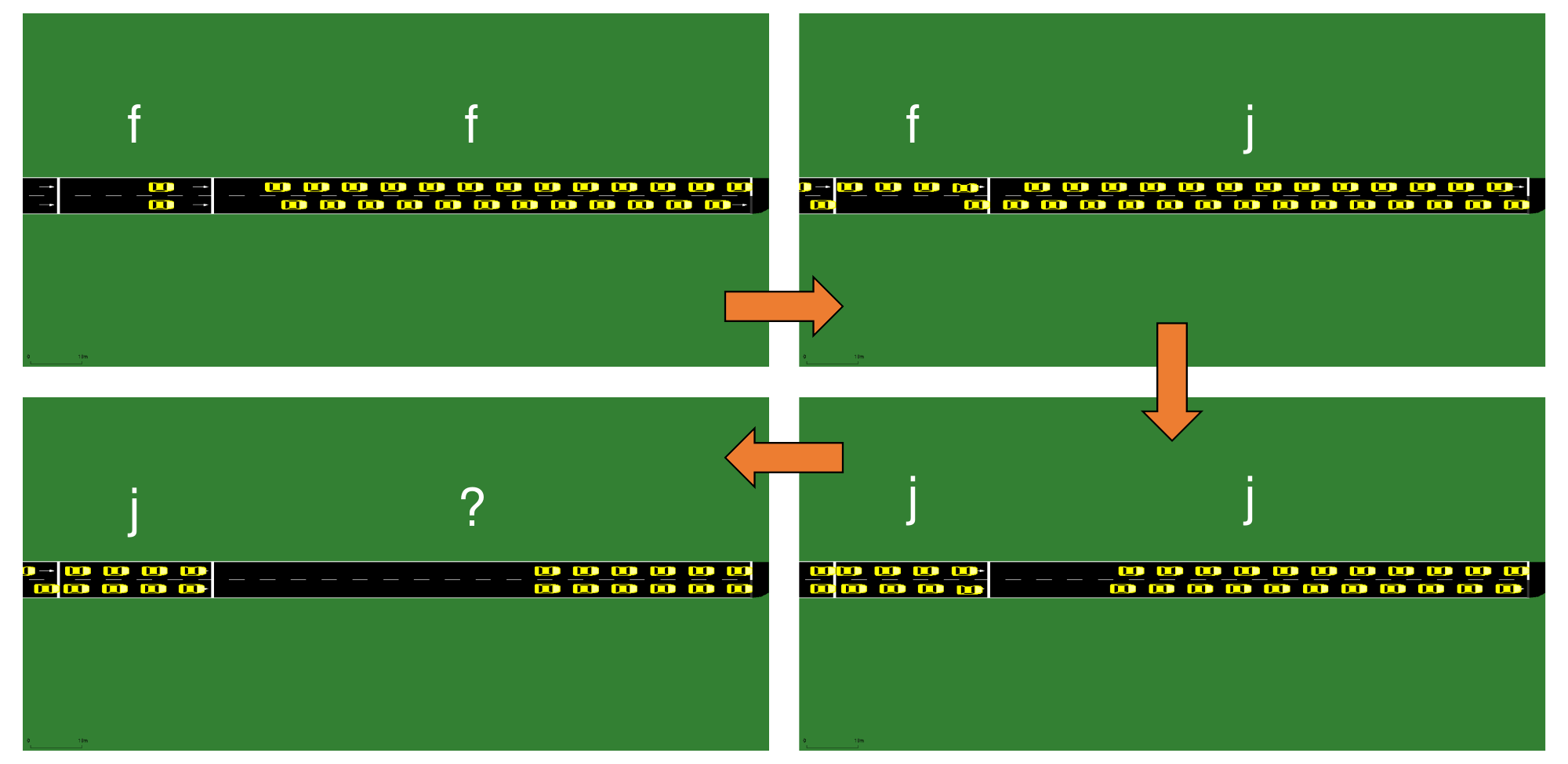}
  \caption{Queue evolution between two edges in SUMO}\label{fig:jj_evol}
\end{figure}

Figure~\ref{fig:jj_evol} shows the queue evolution between two edges in a SUMO simulation. As the queue builds up in the downstream edge, the condition goes from free-free to jam-jam. Due to the late activation of backward traveling space mechanism starting only from the jam-jam condition in the third plot, the density in the downstream edge reaches the maximum for a short while and suddenly drops drastically to a very small number in the fourth plot, while the queue length should ideally remain long during stable congestion.

The headway $\tau_{\text{jf}}$, which aims to capture the capacity drop phenomenon, also does not match the corresponding illustration in Figure~\ref{fig:Eissfeldt}, which actually depicts the condition when a backward traveling space has not yet arrived at the upstream of an edge. Moreover, there was no clear explanation regarding how the traffic condition would move from jam-free to free-free. Overall, the evolution of queue dynamics described in \citet{Eissfeldt2006SimulatingQueues} is problematic.

Without the consideration of backward traveling spaces at all times, the accumulation of queues in the upstream segment is delayed. It is important to emphasize that the distinction of free and jam states is, in fact, an obsolete function in such an LTM-like model.

\subsection{Different cause of traffic congestion} \label{subsec:cause}
\begin{figure}[!ht]
  \centering
  \includegraphics[width=\textwidth]{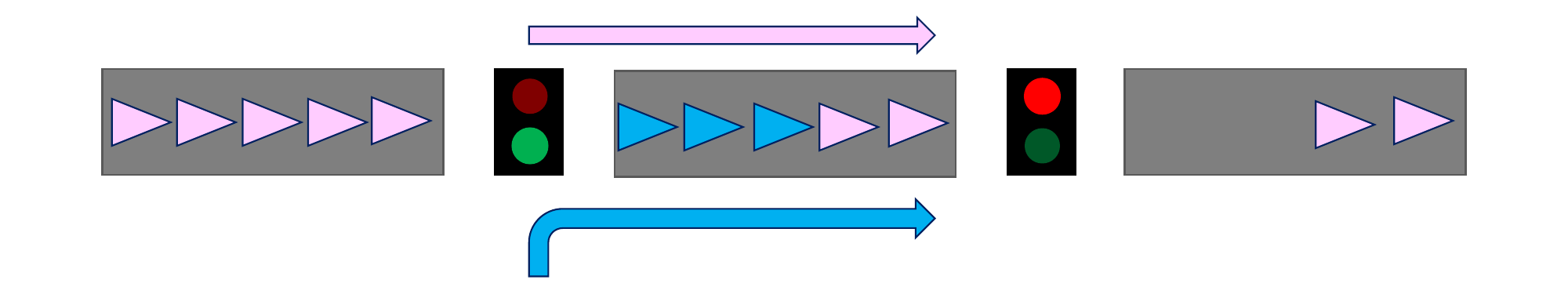}
  \caption{Cause of queue spillback on urban road links (purple: straight-moving vehicles; blue: turning vehicles)}\label{fig:cause}
\end{figure}

The relatively large headway $\tau_{\text{jf}}$ in Eissfeldt's model indicates the reason of congestion growth on motorways. However, it is not the only cause of congestion in reality. In urban road networks for instance, congestion is in fact mainly caused by the imbalanced turning movements \citep{Gayah2011EffectsNetwork} instead of the capacity drop phenomena. As illustrated in Figure~\ref{fig:cause}, several turning vehicles join the queue in the middle segment, making it jammed. Hence, the first vehicle on the upstream segment cannot exit due to the lack of available space, or queue spillback, in the middle segment even when the traffic light is green. The queue length in the upstream segment then starts to increase and affect its upstream segments.

Therefore, to observe the desired congestion propagation pattern, the headway in the jam-free condition $\tau_{\text{jf}}$ does not necessarily need to become larger than the minimum headway $\tau_{\text{ff}}$, especially on urban roads.

The incorporation of capacity drop, which is a second-order traffic dynamics feature, into an LTM-like model that follows the first-order traffic dynamics still needs to be discussed once the distinction of free or jam is removed, as suggested in subsection~\ref{subsec:invalid}.

\section{Mesoscopic link transmission model}
Several studies have developed the mesoscopic and microscopic versions of LTM. The concept can first be found in \citet{Mahut2001ALoading}, even before the macroscopic LTM was proposed, and again in \citet{Mahut2010TrafficDynameq}. Later, \citet{deSouza2025AVehicles} proposed a mesoscopic formulation of LTM that complies with the underlying macroscopic counterpart. 

\begin{figure}[!ht]
  \centering
  \includegraphics[width=\textwidth]{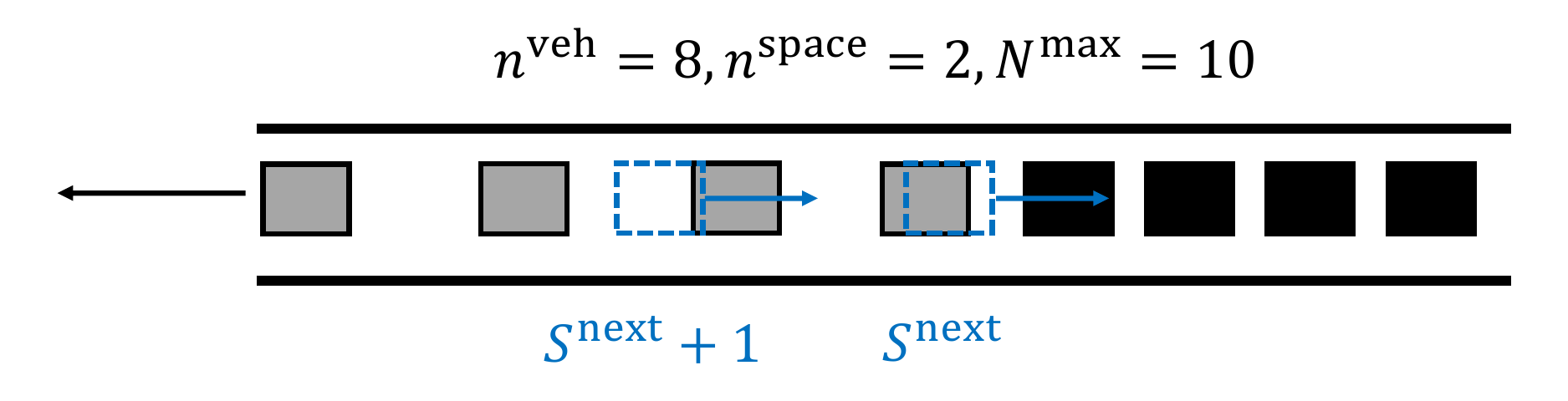}
  \caption{Tracking of the backward traveling spaces (blue dashed rectangles)}\label{fig:space}
\end{figure}

Recently, \citet{Ni2026SimulatingNetworks} proposed LIFT, which is essentially also a mesoscopic LTM but specifically designed for link-level interrupted flow traffic dynamics. This kind of link-based model does not require further discretization of a link and explicitly simulates the movement of backward traveling spaces $S$ to better capture the queue dynamics, as shown in Figure~\ref{fig:space}. One can check if the link has an available space for upstream vehicles to enter by comparing the sum of the number of vehicles $n^{\text{veh}}$ and the number of propagating spaces $n^{\text{space}}$ and the storage capacity of the link $N^{\max}$. The performance of LIFT was validated with both microscopic traffic simulation and empirical data, showing remarkable consistency.

\begin{algorithm}[!ht]
  \caption{Discrete-time LIFT pseudocode}
  \label{alg:lift}
  \begin{algorithmic}[1]

  \State $\texttt{t} \leftarrow 0$

  \While{$\texttt{t} \leq \texttt{T}$}
      \State \textbf{Phase 1: Exit processing}
      \While{\textnormal{eligible exit vehicles exist}}
          \State $\texttt{link*} \leftarrow \displaystyle\arg\min_{\texttt{link}}\ \texttt{link}.\texttt{exit\_supply}$
          \Comment{link with the earliest vehicle exit event}
          \State $\texttt{link*.prev\_exit} \leftarrow \texttt{t}$
          \State $\texttt{link*.space}.\texttt{append(\texttt{t} + \texttt{link*.space\_tt})}$
          \If{$\texttt{link*.veh[0].downstream\_link} \neq \emptyset$}
      \State $\texttt{d\_link} \leftarrow \texttt{link*.veh[0].downstream\_link}$
      \State $\texttt{d\_link.prev\_entry} \leftarrow \texttt{t}$
      \State $\texttt{d\_link.veh.\texttt{append(\texttt{t} +
  \texttt{d\_link.free\_tt})}}$
  \EndIf
          \State delete $\texttt{link*.veh}[0]$
          \Comment{remove exit vehicle from current link queue}
          \State $\textsc{RefreshSupplies}(\texttt{link*},\texttt{d\_link},\ \texttt{t})$
      \EndWhile
      \State \textbf{Phase 2: Entry processing}
      \While{\textnormal{eligible entry vehicles exist}}
          \State $\texttt{link*} \leftarrow \displaystyle\arg\min_{\texttt{link}}\ \texttt{link}.\texttt{entry\_supply}$
          \Comment{link with the earliest entry event}
          \State $\texttt{link*.prev\_entry} \leftarrow \texttt{t}$
          \State $\texttt{link*.veh}.\texttt{append(\texttt{t} + \texttt{link*.free\_tt})}$
          \State delete $\texttt{link*.demand\_queue}[0]$
          \Comment{admit a vehicle from the demand queue}
          \State $\textsc{RefreshSupplies}(\texttt{link*},\ \texttt{t})$
      \EndWhile
      \State $\texttt{t} \leftarrow \texttt{t} + \Delta \texttt{t}$
  \EndWhile

  \end{algorithmic}
  \end{algorithm}

To ensure a fair comparison with MESO, instead of directly using the original continuous-time event-based LIFT, a discrete-time version is developed. Only two headway parameters $\tau_{\text{f}}$ and $\tau_{\text{j}}$, denoting the minimum headway and the time required for the backward traveling space to move a unit vehicle distance, are used in LIFT. Algorithm \ref{alg:lift} explains the modeling logic.

This work uses discrete-time LIFT as an alternative to demonstrate the drawbacks identified in MESO and to show that a mesoscopic model can still generate outcomes that are quite close to microscopic simulation output. 

\section{Case study}
Two scenarios are designed to compare the accuracy of SUMO-MESO and LIFT, a mesoscopic LTM formulation. 

The \texttt{edge-output} function is used to generate aggregated traffic states of every edge from the SUMO simulation.

It is important to mention that there is also no desired speed heterogeneity for validation purposes to completely eliminate the potential influence of stochasticity.

\subsection{Scenario 1}
\begin{figure}[!ht]
  \centering
  \includegraphics[width=\textwidth]{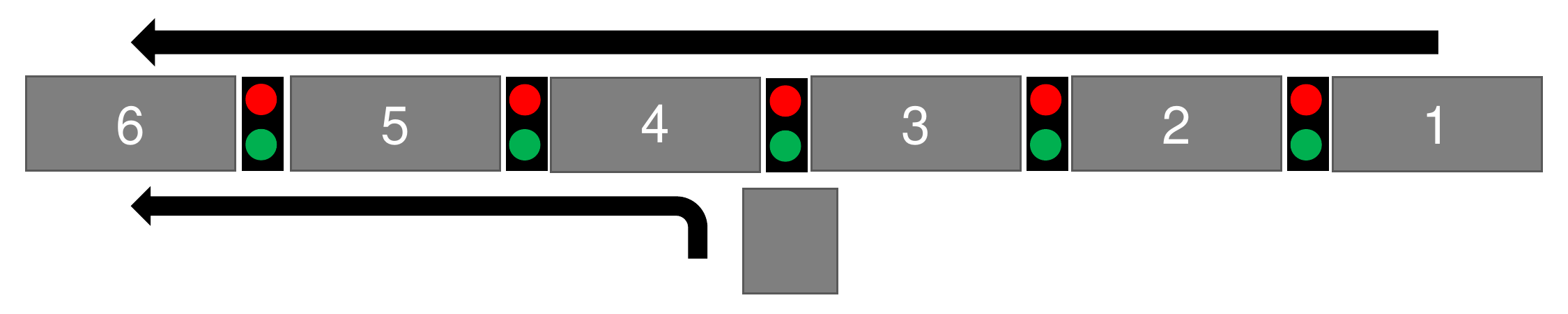}
  \caption{Signalized corridor}\label{fig:corridor}
\end{figure}

A signalized one-way corridor consisting of six 110-m-long single-lane road links with an intersection width of 15 m is designed to evaluate the models, as shown in Figure~\ref{fig:corridor}. The signals along the corridor have a uniform green length of 25 s and a cycle length of 60 s. The signal offset is set to 5 s to introduce a certain degree of signal coordination. 

In addition to the inflow demand that enters from Link 1 and finishes after Link 6, there is another inflow demand from the minor street that joins the mainstream on Link 4. The demand profile gradually increases at the beginning and decreases in the middle of the three-hour simulation period. Consequently, Link 4 becomes the bottleneck and creates spillback that propagates upstream, affecting Links 3, 2, and 1 during the onset of congestion. The congestion then starts to dissipate due to decreasing demand. Therefore, the queue lengths on those congested links would be reduced.

The simulation period is three hours with a time step size of 0.25 s, while the \texttt{actionStepLength} is set to 1.0 s to reflect drivers' decision-making frequency. Table~\ref{tab:param} summarizes the parameter settings. The default car-following model (Krauss) in SUMO is adopted. Preliminary microscopic simulation tests were carried out to calibrate the parameters that are used in the mesoscopic models. In MESO, the same value is used for the headways $\tau_{\text{ff}}$ and $\tau_{\text{jf}}$ to exclude the consideration of capacity drop.

In principle, the speed of a backward traveling space should follow the backward wave speed shown on a triangular FD. However, considering the great unsteadiness of interrupted traffic flow in urban road networks, the value may be slightly faster than the theoretical one. Note that a smaller value leads to faster queue dissipation, while a larger value speeds up congestion propagation. For motorway traffic flow with capacity drop phenomena, a larger value that matches the slope of the congested branch of an inverse-lambda FD may be used when there is a queue to capture the widening congestion wave.

It is worth pointing out that the current MESO is greatly affected by the \texttt{edgelength} setting. It is observed that relatively long values lead to fewer anomalies. Therefore, we select a value that is larger than any edges in the network so that edges are not further segmented.

\begin{table}[!ht]
\caption{List of parameters for the signalized corridor case study}
\label{tab:param}
\centering
        \begin{tabular}[t]{l l l l}
            \hline
            \multicolumn{4}{c}{\textbf{General and MICRO parameters}} \\
            \texttt{desiredMaxSpeed} & 12.5 m/s & \texttt{startupDelay} & 0.75 s \\
            (vehicle) \texttt{length} & 5 m & \texttt{minGap} & 2.5 m \\ \hline
			\multicolumn{4}{c}{\textbf{MESO parameters}} \\
			\texttt{edgelength} & 150 m & \texttt{tauff} $\tau_{\text{ff}}$ & 1.27 s \\
                \texttt{taufj} $\tau_{\text{fj}}$ & 1.27 s & \texttt{taujf} $\tau_{\text{jf}}$ & 1.27 s \\
                \texttt{taujj} $\tau_{\text{jj}}$ & 1.00 s & \texttt{jam-threshold} & -1 \\
                \texttt{junction-control} & true & \texttt{tls-penalty} & 0 s \\ 
                \texttt{multi-queue} & true & \texttt{lane-queue} & true \\ \hline
                \multicolumn{4}{c}{\textbf{LIFT parameters}} \\
                minimum headway $\tau_{\text{f}}$ & 1.67 s & unit backward space travel time $\tau_{\text{j}}$ & 1.00 s \\
                reaction time & 1.0 s & & \\ \hline
		\end{tabular}
\end{table}

\subsection{Scenario 2}

\begin{figure}[!ht]
  \centering
  \includegraphics[width=\textwidth]{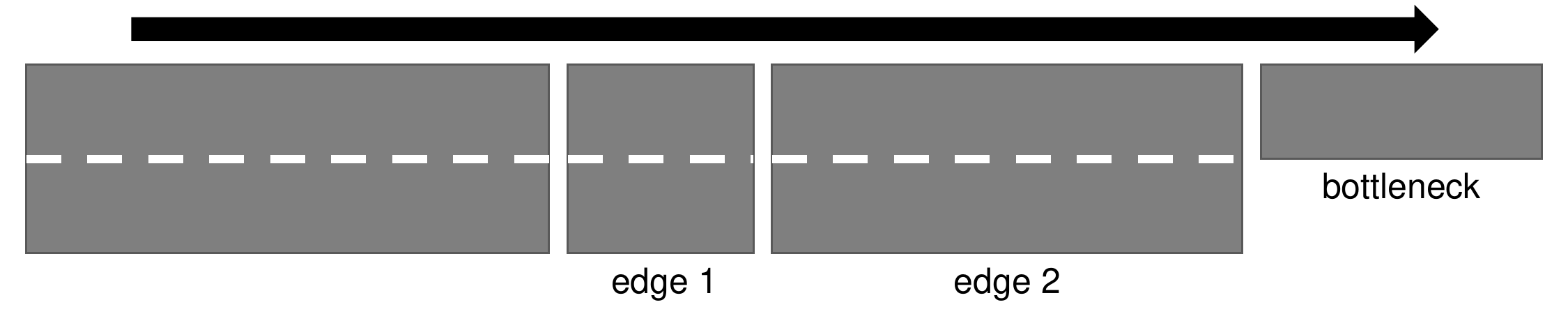}
  \caption{Motorway segment}\label{fig:motorway}
\end{figure}

In the second case study, we aim to test the macroscopic traffic state evolution in a motorway lane-drop bottleneck scenario. A one-hour scenario on a two-lane 250-m long motorway segment with a 50-m long one-lane bottleneck at the downstream is designed, as shown in Figure~\ref{fig:motorway}. The segment is split into three edges. A 105-m long edge (edge 2) is located right before the bottleneck, while a 30-m long edge (edge 1) is further upstream. Another 105-m long edge is implemented only to allow vehicle demand to enter. A higher priority is assigned to the left lane to avoid collisions caused by the merging behavior right before the bottleneck. Parameters for this scenario are summarized in Table~\ref{tab:param2}.

\begin{table}[!ht]
\caption{List of parameters for the motorway segment case study}
\label{tab:param2}
\centering
        \begin{tabular}[t]{l l l l}
            \hline
            \multicolumn{4}{c}{\textbf{General parameters}} \\
            \texttt{desiredMaxSpeed} & 35 m/s & \texttt{startupDelay} & 0.75 s \\
            (vehicle) \texttt{length} & 5 m & \texttt{minGap} & 2.5 m \\ \hline
			\multicolumn{4}{c}{\textbf{MESO parameters}} \\
			\texttt{edgelength} & 150 m & \texttt{tauff} $\tau_{\text{ff}}$ & 1.07 s \\
                \texttt{taufj} $\tau_{\text{fj}}$ & 1.07 s & \texttt{taujf} $\tau_{\text{jf}}$ & 1.07 s \\
                \texttt{taujj} $\tau_{\text{jj}}$ & 1.00 s & \texttt{jam-threshold} & -1 \\
                \texttt{junction-control} & false & \texttt{tls-penalty} & 0 s \\ 
                \texttt{multi-queue} & true & \texttt{lane-queue} & true \\ \hline
                \multicolumn{4}{c}{\textbf{LIFT parameters}} \\
                minimum headway $\tau_{\text{f}}$ & 1.22 s & unit backward space travel time $\tau_{\text{j}}$ & 1.00 s \\ 
                reaction time & 1.0 s & & \\ \hline
		\end{tabular}
\end{table}

The demand profile grows gradually at the beginning and drops immediately after reaching the two-lane capacity. When the inflow demand exceeds the capacity of the one-lane bottleneck, congestion occurs and propagates upstream throughout the motorway segment.

The microscopic simulation counterpart of this scenario is not available as it is challenging to calibrate the lane-changing behavior to match the theoretical macroscopic traffic flow phenomena. This is considered an interesting future work direction.

\section{Results}
\begin{figure}[!ht]
  \centering
  \includegraphics[width=\textwidth]{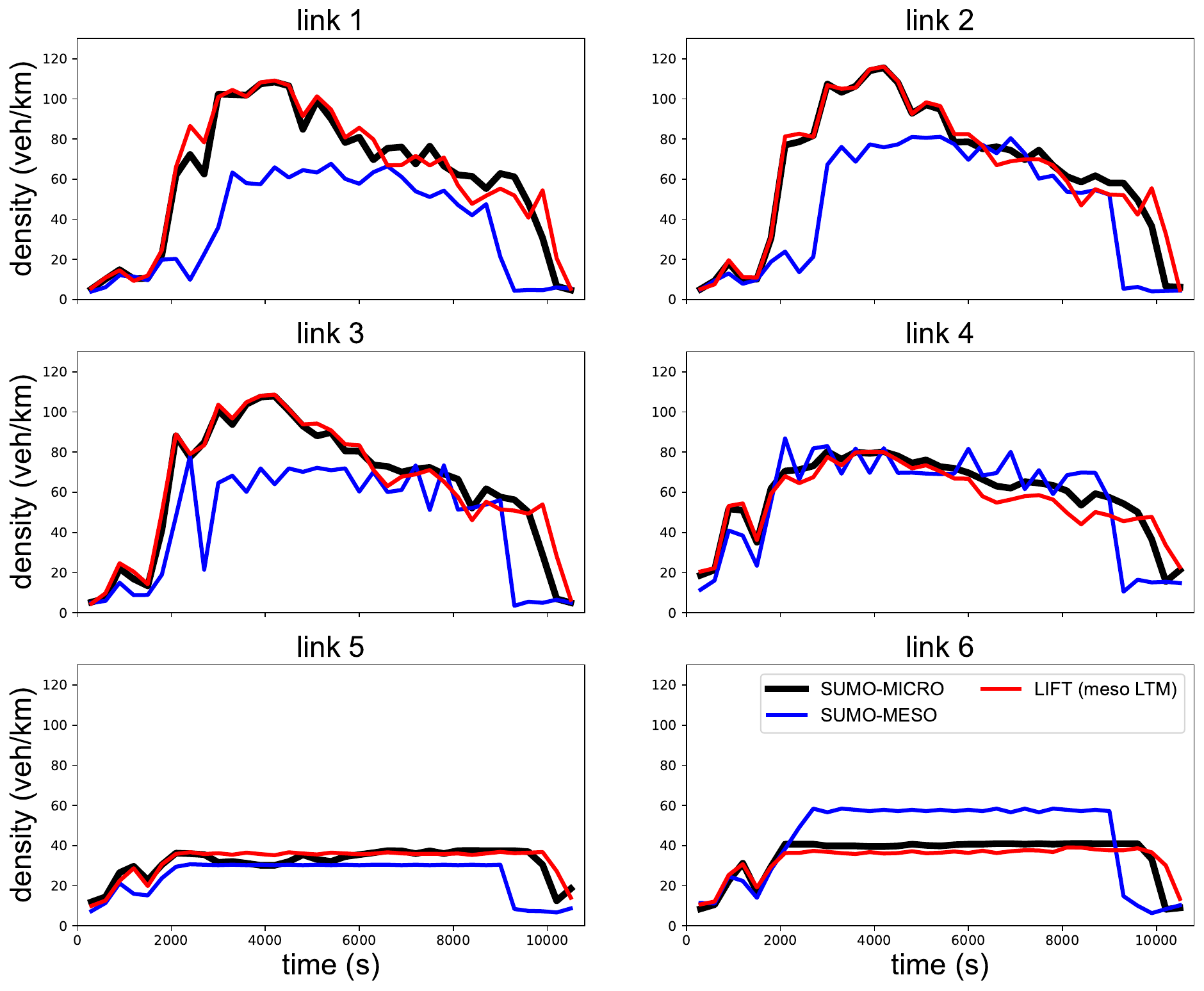}
  \caption{Link density evolutions on the signalized corridor}\label{fig:density}
\end{figure}

The evolution of link densities, which is an indicator of macroscopic link traffic state, in the first case study scenario is analyzed with an aggregation period of five minutes. Figure~\ref{fig:density} shows the simulation results from every model.

The simulation output produced by MESO shows underestimated link densities during the congestion at the upstream of the bottleneck on links 1--3, while congestion also occurred later and dissipated earlier on these links. This demonstrates the influence of the invalid queue dynamics, as explained in subsection~\ref{subsec:invalid}.

In contrast, the results produced by LIFT are consistent with the microscopic simulation output. The onset of congestion and the gradual dissipation of congestion are captured successfully.

For the motorway scenario, the one-minute aggregated traffic states of the two edges are shown on the flow-density plane and compared with the theoretical FD inferred following the parameters. 

\begin{figure}[!ht]
  \centering
  \includegraphics[width=\textwidth]{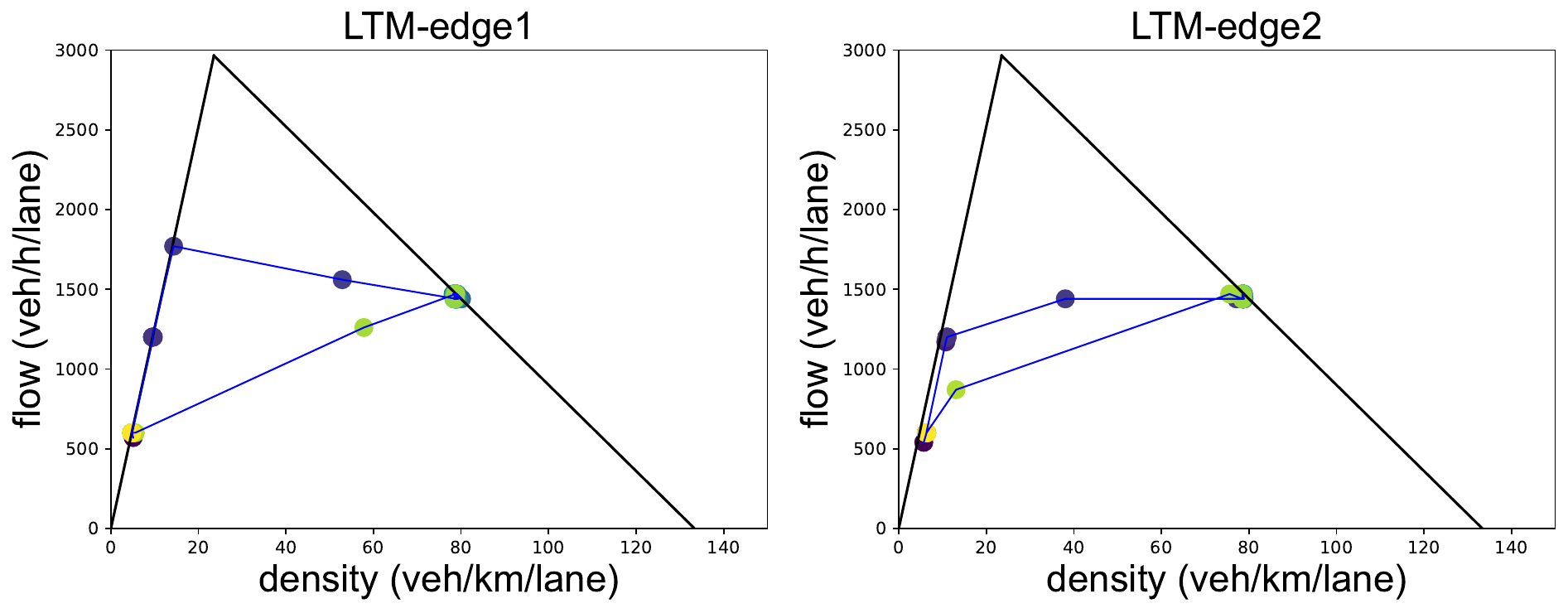}
  \caption{Macroscopic traffic states of the motorway segments produced by LIFT (meso LTM)}\label{fig:FD_LTM}
\end{figure}

\begin{figure}[!ht]
  \centering
  \includegraphics[width=\textwidth]{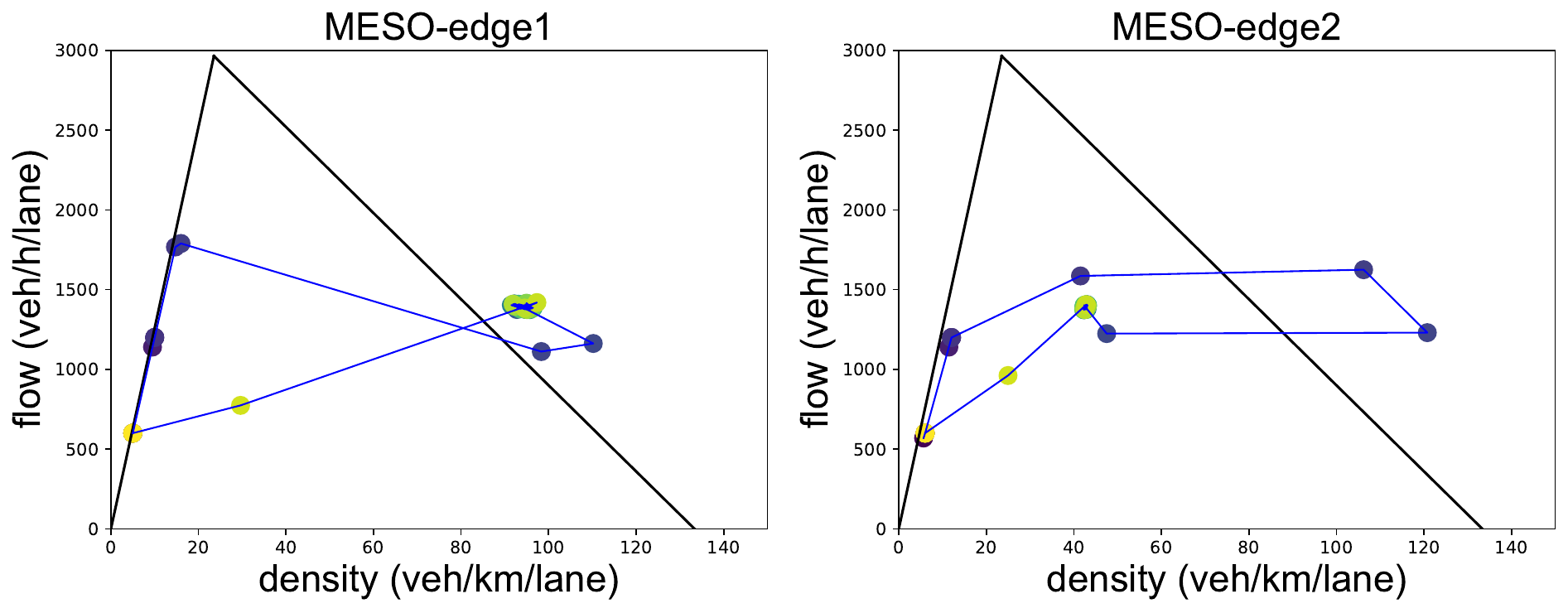}
  \caption{Macroscopic traffic states of the motorway segments produced by MESO}\label{fig:FD_MESO}
\end{figure}

As shown in Figure~\ref{fig:FD_LTM}, the traffic states produced by LIFT on both edges follow the LWR kinematic wave theory analysis and stay close to the FD envelope except a few short periods during the transition between free-flow and congested regimes, where the link-internal traffic heterogeneity is large. However, Figure~\ref{fig:FD_MESO} shows that the macroscopic traffic states produced by MESO go beyond the FD and contain larger state heterogeneity within the edges, particularly the long edge close to the bottleneck.

\section{Conclusions}

Unlike macroscopic or microscopic models, the theoretical validity of many mesoscopic traffic flow models has not been carefully examined. This is likely because of the limited understanding of macroscopic traffic flow theory in the past. The simplified traffic dynamics can significantly affect the accuracy of simulation outcomes.

In view of this problem, this paper identifies deficiencies in Eissfeldt's model and the current SUMO-MESO function and demonstrates their influence in two case study scenarios by analyzing the macroscopic traffic state evolutions. The current SUMO-MESO is found to systematically underestimate congestion.

To address the drawbacks, an alternative that closely follows the macroscopic LTM is proposed. The case study results show that a carefully-designed mesoscopic traffic flow model following the principle of traffic flow theory can yield results that are highly consistent with kinematic wave theory and microscopic traffic simulation. Future work hopes to integrate it into the SUMO source code.

\section*{Declaration of competing interest}
The authors declare that they have no known competing financial interests or personal relationships that could have appeared to influence the work reported in this paper.

\section*{Acknowledgement}
The authors would like to thank Dr. Lukas Amb{\"u}hl, Dr. Matteo Felder, and Dr. Sasan Amini at Transcality AG and Dr. rer. nat. Jakob Erdmann at the SUMO team in Deutsche Zentrum f{\"u}r Luft- und Raumfahrt for the helpful discussion.

\section*{Data availability}

The Python code for discrete-event LIFT can be accessed here \url{https://github.com/YCNi/LIFT.git}.

The SUMO simulation packages of the two case study scenarios can be found here \url{https://github.com/YCNi/SUMO_MESO_CASES.git}.

It is worth mentioning that the SUMO team is actively implementing its meso-LTM function with the help from the authors at the moment when this article is published.

\bibliographystyle{model5-names}
\bibliography{references}
\end{document}